\documentclass[11pt]{article}
\usepackage{array}
\usepackage{graphicx}
\usepackage{textpos}

\title{Emergent Majorana Mass and Axion Couplings in Superfluids}

\author{Frank Wilczek\\
\small\it Center for Theoretical Physics, MIT, Cambridge MA 02139 USA}

\begin{document}

\maketitle

\begin{textblock*}{5cm}(11cm,-8.2cm)
  \fbox{\footnotesize MIT-CTP-4529}
\end{textblock*}

\begin{abstract}
Axions (in the general sense) may acquire qualitatively new couplings inside superfluids.   Their conventional couplings to fermions, in empty space, involve purely imaginary masses; the new couplings involve emergent Majorana masses.  The possibility of weak links for axions, recently put forward, is analyzed and replaced with a non-local analogue.
\end{abstract}

\medskip

\bigskip

Interactions in the gauge sector of the standard model are powerfully constrained by general principles of quantum field theory and symmetry, as is its interface with general relativity.   In the flavor sector, where fermion masses and mixing arise, known symmetries have much less power, and theoretically unconditioned parameters proliferate.   Two promising, though as yet hypothetical, ideas could explain striking qualitative features within that sector.  One is that the flavor sector supports hidden symmetries, that are broken only spontaneously or by quantum anomalies (or both).  An especially compelling case is Peccei-Quinn (PQ) symmetry \cite{PQ, PQReview}, which could explain the otherwise mysteriously tiny value of the phase of the determinant of the quark mass matrix, or equivalently the effective $\theta$ parameter of quantum chromodynamics (QCD).    If some flavor symmetries are continuous and spontaneously broken, they lead to a characteristic phenomenological consequence: the existence of very light spin-0 particles, whose properties are closely connected with broken symmetry \cite{familons}.   We will call such particles axions, following a generalization of the original usage that is now very common.   Another, which applies to neutrinos, is that their masses may be of a special type: Majorana masses \cite{Majorana, MajoranaReview}.   That possibility is favored in unified field theories, and in that context it is connected to the otherwise mysteriously tiny scale of neutrino masses.  

Here I will demonstrate a conceptual connection between those two ideas, that arises in the analysis of axion couplings in superfluids.   That subject is interesting, of course, in guiding the continuing experimental search for axions.   The analysis also sheds, I think, considerable light on the nature of Majorana mass and Majorana fermions.   Recently a possible ``weak link'' coupling of axions to superconductors was suggested \cite{beck}.   Although I do not agree with that suggestion, for reasons discussed below, it stimulated the work reported here.  

\bigskip

\section*{Axion Vector Current Coupling}

In general, axions will be spin-0 bosons coupled to the divergence of a symmetry-breaking current.  That is an abstract, generalized form of the Goldberger-Treiman relation \cite{goldbergerTreiman, GTReview}.   For definiteness, and because it illustrates the main points in a transparent form, let us consider a symmetry that acts on both right- and left-handed electrons, with charges $b, c$ respectively.   (We have in mind that our symmetry may be broken spontaneously well above the weak scale, so that this distinction is relevant.   The model of the following Section will embody this framework concretely.)   Thus the symmetry current has both vector and axial vector pieces:
\begin{equation}
j^\mu ~=~ b \, \overline{e_R} \gamma^\mu e_R \, + \, c  \, \overline{e_L} \gamma^\mu e_L \, ~=~ \frac{b+c}{2} \bar e \gamma^\mu e \, + \, \frac{b-c}{2} \bar e \gamma^\mu \gamma_5 e
\end{equation}
The vector piece is usually neglected, because its divergence (usually) vanishes; indeed, electron number is (usually) conserved.   In a superconductor, however, when one expands around the stable ground state, electron number is not conserved in the usual ``bookkeeping'' (Wigner-Weyl) sense.  Indeed, there is a non-trivial condensate, and which effectively renders electron number indefinite.  More precisely: The conservation law associated with a spontaneously broken symmetry is realized in the Nambu-Goldstone mode, with cancellation between the divergence of the current and the singular contribution from coupling to a light boson.   In our context, the divergence of the electron number current can induce additional axion couplings.

\subsection*{Analogy of Superconducting Gap and Majorana Mass}

To see it, consider the effective coupling of electrons to the condensate, which represents the electron number violation.  Suppressing spin indices, and considering only simple s-wave ordering, we have the effective interaction
\begin{equation}\label{gapInteraction}
{\cal L}_{\rm electron-condensate} ~=~ \Delta^* \, e e \, + \, {\rm h.c.} ~\leftarrow~ \kappa \, \bar e \bar e e e + {\rm h.c.}
\end{equation}
arising from the condensation $\Delta = \kappa \langle ee \rangle$.   Famously, this interaction opens a gap in the electron spectrum at the fermi surface.   

A close analogy between the opening of this gap and the generation of mass, by condensation, for relativistic fermions was already noted in the earliest work on spontaneously broken symmetry in relativistic particle physics, and indeed largely inspired that work.   Revisiting that analogy, we discover its relevance to a basic issue in contemporary physics: the question of {\it Majorana mass\/} for neutrinos, which I now briefly recast into a form suggestive for our purpose.   

Neutrino oscillations provide evidence for mass terms that are not diagonal with respect to the separate lepton numbers (though as yet no observation has revealed violation of the total $L_e + L_\mu + L_\tau$).   Mass terms, diagonal or not, are incompatible with chiral projections.  Thus the familiar ``left-handed neutrino'', which particle physicists worked with for decades, can only be an approximation to reality.   The physical neutrino must have some admixture of right-handed chirality.  Thereby a fundamental question arises: Are the right-handed components of neutrinos something entirely new -- or could they involve the same degrees of freedom we met before, in antineutrinos?  At first hearing that question might sound quite strange, since neutrinos and anti-neutrinos have quite different properties.   How could there be a piece of the neutrino, that acts like an antineutrino?   But of course if the piece is small enough, it can be compatible with observations.  Quantitatively: If the energy of our neutrinos is large compared to their mass, the admixture of opposite chirality will be proportional to $m/E$.   To explain the phenomenology of neutrino oscillations, and taking into account cosmological constraints, we are led to masses $m < {\rm eV}$, and so in most practical experiments $m/E$ is a very small parameter.  

So: Are neutrinos and antineutrinos the same particles, just observed in different states of motion?  The observed distinctions might just represent unusual spin-dependent (or more properly helicity-dependent) interactions.   These questions are usually posed in the cryptic form: Are neutrinos Majorana particles?  

To pose the questions mathematically, we must  describe a massive spin-$\frac{1}{2}$ particle using just two (not four) degrees of freedom. We want the antiparticle to involve the same degrees of freedom as the particle. Concretely, we want to investigate how the hypothesis
\begin{equation}\label{MajoranaHypothesis}
\psi_R ~\stackrel{?}{=}~ \psi_L^{\ *}
\end{equation}
(in a Majorana basis, with all $\gamma^\mu$ matrices pure imaginary) might be compatible with non-zero mass.  Applying a chiral projection to the Dirac equation in general gives us the form
\begin{equation}
i \gamma^\mu \partial_\mu \psi_L + M \psi_R ~=~ 0
\end{equation}
and so we are led to contemplate
\begin{equation}\label{MajoranaEquation}
i \gamma^\mu \partial_\mu \psi_L + M \psi_L^{\ *} ~=~ 0
\end{equation}

(Mathematical/historical aside: If Eqn.\,(\ref{MajoranaHypothesis}) holds, we can derive both $\psi_L$ and $\psi_R$ by projection from a single four-component {\it real\/} field, i.e.  
\begin{equation}
\psi ~\equiv~ \psi_L + \psi_R ~=~ \psi_L + \psi_L^{\ *}
\end{equation}
This is the link to Majorana's original concept of a real spin-$\frac{1}{2}$ field.) 

The appearance of Eqn.\,(\ref{MajoranaEquation}) is unusual, and we may wonder how it could arise as a field equation, from a Lagrangian density.   Usually we consider mass terms
\begin{equation}
{\cal L}_M ~\propto~ \bar \psi \psi ~=~ \psi^\dagger \gamma_0 \psi
\end{equation}
Now if we write everything in terms of $\psi_L$, using Eqn.\,(\ref{MajoranaHypothesis}), we find
\begin{equation}\label{MajoranaMassTerm}
{\cal L}_M ~\propto~ \psi^\dagger \gamma_0 \psi ~\rightarrow~ (\psi_L)^T \gamma_0 \psi_L + {(\psi_L^{\ *})}^T \gamma_0 \psi_L^{\ *}
\end{equation}
where $^T$ denotes transpose.  In verifying that these terms are non-trivial, whereas the remaining cross-terms vanish, it is important to note that $\gamma_5$ is antisymmetric, i.e., that it changes sign under transpose.   That is true, because $\gamma_5$ is both Hermitean and pure imaginary.   Varying this form, together with the conventional kinetic term 
\begin{equation}
{\cal L} ~\propto~ {(\psi_L^{\ *})}^T \, i\gamma_0 \gamma^\mu \partial_\mu \psi_L + h.c.
\end{equation}
will give us Eqn.\,(\ref{MajoranaEquation}).  

A close analogy between the Majorana mass term Eqn.\,(\ref{MajoranaMassTerm}) and the gap-opening interaction Eqn.\,(\ref{gapInteraction}) is evident.   Both are number-violating, derivative-free quadratic terms.  Their physical consequences are also closely analogous.   Electron quasi-particles near the fermi surface in a superconductor are their own antiparticles, in an evident sense: a pair of quasi-particles with equal and opposite momenta $\pm k$ (and spins) has vacuum quantum numbers, since their superposition overlaps with the condensate.    Inside superconductors, electrons are Majorana fermions, in this broad sense.   (In several other, more special, situations there is a closer approach to relativistic kinematics \cite{solidMajorana}.   The excitations associated with localized Majorana modes  \cite{kitaevWire, majoranaMode}, or ``Majorinos'' \cite{majorinos}, are remarkable objects that can be considered as massless Majorana particles in space-time dimensions 0+1 -- i.e., zero energy excitations supported on points).  

It should be noted that particles, whether electrons in superconductors, neutrinos, or even (as discussed below) scalars can support both Majorana and ordinary, number-conserving masses.   One can make a continuous interpolation between ``Majorana'' and ``conventional'' particles.  For electrons, specifically, the Majorana mass dominates only in a small kinematic region near the nominal Fermi surface in a superconductor.  

\subsection*{Result and Model}

Returning to the axion coupling, we find that the divergence of the vector current gives us an axion coupling
\begin{equation}\label{aSuper}
{\cal L}_{\rm a-super} ~=~ -  i\frac{a}{F} (b+c) \ \Delta^* \, e e +  {\rm h.c.})
\end{equation}
(This is the small-field form; in general fwe should replace $i\frac{a}{F} \rightarrow e^{i\frac{a}{F}}$, understanding that ordinary gap mass term is included.)
This can be compared to the usual ``vacuum'' coupling, which arises entirely from the divergence of the axial vector current
\begin{equation}\label{aVacuum}
{\cal L}_{\rm a-vac} ~=~ - i\frac{a}{F} (b-c) \ m_e \bar e \gamma_5 e
\end{equation}
In the non-relativistic limit, this represents a momentum- and spin-dependent interaction.   (It still contributes inside a superconductor, of course.)
We can summarize the situation by saying that Eqn.\,(\ref{aVacuum}) gives a coupling to an {\it imaginary\/} mass of magnitude $m_e$, while Eqn.\,(\ref{aSuper}) gives a coupling to a {\it Majorana\/} mass of magnitude $| \Delta |$.

In this section I outline the construction of a simple microscopic model that embodies this concept, with $c=0$.   It is, in fact, essentially the original axion model of \cite{wilczek, weinberg}, modified to allow the possibility of a large (compared to electroweak) PQ symmetry breaking scale \cite{invisible1}.  (Alternative axion schemes \cite{invisible2}, where all the action is in the hadronic sector, have $b=c=0$ for electrons.)   

We contemplate a model with $U(1)_{\rm local}\times U(1)_{\rm local} \times U(1)_{\rm global}$ symmetry, meant to be interpreted as incorporating a truncation of the standardl model, containing three complex scalar fields $\phi, \phi_1, \phi_2$, and of course electrons of two chiralities $e_L, e_R$.   $\phi_1$ and $\phi_2$ represent the upper, electrically neutral components of two Higgs doublets, and the first $U(1)$ implements phase transformations on them and on $e_R$:
\begin{eqnarray}
(\phi_1, \phi_2) ~&\rightarrow&~ e^{i\alpha} \, (\phi_1, \phi_2) \nonumber \\
e_R ~&\rightarrow&~ e^{-i\alpha} \, e_R \nonumber \\
(\phi, e_L) ~&\rightarrow&~ (\phi, e_L)
\end{eqnarray}
The second $U(1)$ is electromagnetism, which acts as
\begin{eqnarray}
(\phi, \phi_1, \phi_2) ~&\rightarrow&~ (\phi, \phi_1, \phi_2) \nonumber \\
(e_R, e_L) ~&\rightarrow&~ e^{i\beta} (e_R, e_L)
\end{eqnarray}
The third $U(1)$ is PQ symmetry, which acts as 
\begin{eqnarray}\label{PQTransformation}
(\phi, \phi_1, e_R) ~&\rightarrow&~ e^{i\gamma} (\phi, \phi_1, e_R) \nonumber \\
\phi_2 ~&\rightarrow&~ e^{-i\gamma} \phi_2 \nonumber \\
e_L ~&\rightarrow&~ e_L
\end{eqnarray}

Now we suppose that $\phi_1, \phi_2$ acquire vacuum expectation values $v_1, v_2$ at the electroweak scale, while $\phi$ acquires a much larger vacuum expectation value $F$.    Then the soft mode associated with smooth space-time variation in $\alpha$ gets ``eaten'', according to the Higgs mechanism, while we get a physical soft mode associated with smooth space-time variation in $\gamma$.  Electromagnetic $U(1)$ is unbroken by these condensations of neutral fields.  The physical soft mode is generated by acting with Eqn.\,(\ref{PQTransformation}) with a space-time dependent $\gamma$.   The quanta of this soft mode are axions.  Linearizing around the condensates, we find that the axion field, normalized to have canonical kinetic energy, is  
\begin{equation}\label{axionField}
a ~=~ \frac{F {\rm Im} \phi  + v_1 {\rm Im} \phi_1 - v_2 {\rm Im} \phi_2}{\sqrt{F^2 + v_1^2 +  v_2^2} }~\approx~ {\rm Im} \phi  + \frac{v_1}{F} {\rm Im} \phi_1 - \frac{v_2 }{F} {\rm Im} \phi_2
\end{equation}
Because the axion is a soft mode associated to a PQ symmetry Eqn.\,(\ref{PQTransformation}) that only moves the right-handed electron, we find that this microscopic model realizes the framework sketched previously, with $b=1, c=0$.


\section*{Comments}

\begin{enumerate}
\item Both couplings Eqns.\,(\ref{aSuper}, \ref{aVacuum}) support the possibility of exciting electron pairs over the gap with a time dependent axion field, such as might be responsible for the astronomical dark matter.  In that context, the frequency dependence is essentially $a \propto e^{-im_at}$, where $m_a$ is the axion mass.  In particle language, one has the absorption process $a \rightarrow ee$.   The coupling Eqn.\,(\ref{aSuper}), with its simple form, might also support more delicate effects, that depend on quantum coherence (as might the spin-dependent coupling, for spin-dependent condensates).  These possibilities deserve further study.
\item Similar considerations apply to axions of other types, and their couplings to other sorts of superfluids, such as liquid $^4$He, or possible hadronic condensates in neutron stars.   In the latter application, of course, much larger gap sizes are in play.
\item It is instructive to consider the analogue of ``Majoranization'' through mass acquisition, for bosons.   If we have a global $U(1)$ symmetry 
\begin{equation}
(\phi, \phi_1) ~\rightarrow~ e^{i\alpha} (\phi, \phi_1)
\end{equation}
broken by $\langle \phi \rangle = v $ condensation, then mass terms arising from 
\begin{eqnarray}
{\cal L}_m ~&=&~ -\kappa \, \phi^{2*} \phi_1^2 \, + \, {\rm h.c.} ~\rightarrow~  -\kappa v^2 (\phi_1^2 + \phi_1^{*2}) \nonumber \\
~&=&~ -2\kappa v^2 \bigl( ({\rm Re} \phi_1)^2 - ({\rm Im} \phi_1)^2 \bigr) 
\end{eqnarray}
will split the quanta produced by the real and imaginary parts of $\phi_1$, and thus tend to lift the degeneracy of quanta that had opposite $U(1)$ charge, and formed particle-antiparticle pairs, in the unbroken symmetry state.   
\item Formally, any non-zero Majorana mass splits the underlying charged state into two neutral ``Majorana fermions'', but since that splitting can be arbitrarily small, the binary distinction (i.e., Majorana versus Dirac) can be misleading.  For example, if neutron-antineutron oscillations are allowed, the true mass eigenstates are Majorana `neuterons', which are very nearly coherent superpositions $\frac{1}{\sqrt 2} ( | n \rangle \pm e^{i \phi} | \bar n \rangle )$ of neutron and antineutron.  For practical purposes, however, the important states are the pure neutron and antineutron states, since strong interactions cause the neuterons to decohere.   Similarly, the Majorana mass terms for electrons in superconductors only dominate their behavior for momenta in a small range near the nominal Fermi surface, as mentioned previously.  In that context, however, it has direct physical implications \cite{beenakker}. 
\item It is possible that the right-handed neutrino $N_R$, which figures in the see-saw mechanism for light neutrino (Majorana) mass generation, has non-trivial Peccei-Quinn charge, and that its mass arises directly from its coupling to $\phi$, in the form
\begin{equation}
{\cal L}_M ~=~ -M \bigl( N_R^T \gamma_0 N_R + {N_R^{\ *}}^T \gamma_0 N_R^{\ *} \bigr)~\stackrel{\propto}{\leftarrow}~ \kappa \phi^2  N_R^T \gamma_0 N_R + {\rm h.c.}
\end{equation}
This would lead to a substantial axion coupling $\propto M/F$, which might have cosmological implications.  
\item There are no constructible weak links in PQ symmetry
breaking.  That symmetry breaking, which occurs at an enormous energy scale, is universal and robust, quite unlike the symmetry breaking of
superconductivity.  The magnitude of the superconducting order parameter is material dependent and can be made very small at
Josephson junctions, and effectively zero outside material circuits, whereas the magnitude of the PQ condensate (i.e., $F$) is enormously large, compared to practical experimental energy scales.   As a mathematical consequence, the axion field is
single-valued, so one should put the integral of its derivative around a loop equal to zero.   Indeed, for there to be an integrated phase, the absolute value of the underlying order parameter field must
vanish somewhere inside the loop, as it does in the core of a cosmic axion string.  Thus the key equation (Equation 3) of \cite{beck}, which sets up a relation between between the axion field (regarded as a phase) integrated around a loop, and the corresponding quantity for the superconducting phase, reduces vacuously to the usual Josephson circuit equation, with no axion contribution.  Addition of the axion term in any case had no apparent physical basis, since the axion field, unlike the superconducting phase field, is invariant under electromagnetic gauge transformations, contrary to Equation 5 of \cite{beck}.
\item A related point is that variations of the hypothetical cosmic background axion field are expected to be very small, both in space and in time, on scales relevant to ordinary Josephson junctions.  Insofar as it does not vary, it is essentially equivalent to a redefinition of the phase of the electron field (the overall, not any relative, phase) -- which is unobservable.   It may be interesting to consider non-local Josephson effects \cite{iazzi, beenakker} in this regard.  
\item To reconcile the preceding point with the basic coupling Eqn.\,(\ref{aSuper}), we must consider an important limitation of that formulation, as applied to finite bodies.  The integration by parts that allows us to re-write the primary, gradient coupling $\partial_\mu a j^\mu$ as $-a \partial_\mu j^\mu$ must be done with care, taking into account that condensates vanishes outside the bodies which support them.   That leads to surface terms.  In fact those surface terms cancel off the whole answer for very long-wavelength variations in $a$, as they must since (for example) electron number is ultimately conserved. \end{enumerate}

In summary, I have demonstrated the existence of a new form for possible coupling of axions, that in particular arises within superfluids.  These couplings respond to emergent as well as fundamental Majorana mass terms.  Several possible phenomenological applications have been mentioned, but their quantitative consideration is left to future work.

\bigskip

\subsection*{Acknowledgement} I wish to thank Matthew McCullough for helpful conversations on several aspects of this work, and Alex Sushkov and Lawrence Krauss for discussions on nonlocal Josephson effects.  This work was supported by the U.S. Department of Energy under contract No. DE-FG02-05ER41360.

\end{document}